\documentstyle[preprint,prl,aps,epsf]{revtex}

\begin{document}

\draft

\preprint{\vbox{\hbox {December 1996}\hbox {rev. Jan. 1997} \hbox{IFP-726-UNC}}}

%@twocolumnfalse\endcsname

\title{Strong CP and Low-energy Supersymmetry}
%\title{How to Define the Minimal Supersymmetric Aspon Model}

\author{\bf Paul H. Frampton and Otto C. W. Kong}
\address{
Institute of Field Physics, Department of Physics and Astronomy,\\
University of North Carolina, Chapel Hill, NC  27599-3255}
\maketitle
%\date{\today}

\begin{abstract}
A spontaneously-broken CP provides an alternative to the KM mechanism 
for CP violation with the advantage that the strong CP problem is solved. 
We consider, for such a model with a new gauged $U(1)$, 
the incorporation of low-energy supersymmetry and find
the constraints on alignment and squark degeneracy. The conclusion is that
although the $\bar{\theta}$ constraints are much less severe than in other generic schemes with supersymmetry breaking and spontaneous CP
violation, one restriction remains stronger than needed in the
MSSM for suppression of FCNC.

\end{abstract}

%\vskip0pc]
%\vskip2pc]

\newpage 
%\narrowtext

Symmetries play a fundamental role in physics; in particular, study of
discrete spcaetime symmetries like P and T have revolutionized 
our theory of particle physics during the last forty years. Our present
understanding of P violation is incoporated as a part of the standard
model in the form of chiral fermions. Our view of T (or equivalently CP)
violation is less mature and requires the acquisition of more empirical data.

A model of spontaneous CP violation (SCPV) with an extra gauged
$U(1)$ symmetry  was first proposed\cite{aspon1} in 1990
and developed in subsequent papers\cite{aspon2,aspon5}. The
principal advantage over the standard model is that the
strong CP problem is solved. The gauged $U(1)$ also provides a new
mechanism for generation CP-violating effects in neutral meson mixings.
In the recent work with  Glashow\cite{aspon5}, it was emphasized how 
the (aspon) model is fully consistent with present experimental data and that a
testable prediction is made in $B$-decay (see also \cite{ab}).

Here we consider whether the essence of the aspon mechanism can
co-exist with low-energy supersymmetry (SUSY). In particular, we address
the question of what  the minimal supersymmetric aspon model
(MSAM) is. An important requirement is that the MSAM permits SCPV
in its Higgs potential. Also, we wish to specify the constraints
on the soft SUSY breaking parameters (SSBP), e.g. 
proportionality of $A$-terms  and squark mass degeneracy,
which must be satisfied for consistency with experiment. Our aim here
is not to make any specific proposal about how such constraints 
may be satisfied, though we will discuss the part of the issue related
to specifying  fully the MSAM and make some speculations beyond that.
We hope to return to the question in  future publications.

There already exists a considerable literature on the
question of SCPV in supersymmetric extensions of the standard
model, so we need to explain how the present paper differs
from earlier work. It is well known that  SCPV is not possible in the
tree-level Higgs potential of a supersymmetric standard model with
minimal Higgs content. The papers\cite{romao,masip}
study some alternative possibilities and arrive at interesting
no-go theorems which rule out certain interesting classes of extended Higgs
sectors. A model with an extra pure singlet Higgs, however, admits
SCPV\cite{pomarol2}. We shall use these results in defining our MSAM.
The work of \cite{dine} gives constraints on the proportionality
and degeneracy necessary for phenomenological consistency
in generic models with SUSY and an SCPV solution to the strong
CP problem (see also\cite{bar}). The authors
then conclude that the constraints on the SSBP are much more severe than
the corresponding ones from FCNC and cannot expected to be 
satisfied without unnatural fine tunings.
However, in \cite{dine} the additional
quark is assumed to have very heavy $\geq 10^{11}$GeV
mass while in the aspon case discussed here, the new quark(s)
are instead expected to be relatively light, below 600GeV\cite{aspon2}
for example. A second difference from \cite{dine} is that the aspon
model provides an additional mechanism for CP violation in the kaon
system and so the constraint provided by the only measured
CP violation parameter $\epsilon$ is quite different.

The fields of the non-supersymmetric aspon model comprise
the standard model with three families, together with
a vector-like doublet of quarks $Q_o$, two complex
scalar singlets $\chi_{1,2}$ and the gauge field (aspon) of an additional
$U(1)_{a}$ with respect to which only the extra quarks and scalars
are charged. The first question then is whether the simplest possible
MSAM is to take just the same fields rewritten as superfields?
To cancel anomalies of the fermionic partners $\tilde{\chi}_{1,2}$
we must introduce the conjugate superfields, designated $\chi_{3,4}$.
The latter have no admissibleYukawa couplings to the quark superfields.
But even then one must ensure that $\left\langle\chi_{1,2}\right\rangle$ can be complex
as necessary for the aspon scenario?  

At {\it tree} level the resultant Higgs potential is sufficiently 
similar to that discussed in \cite{masip} that we deduce that
SCPV can occur only at {\it isolated} points in parameter space
and is therefore unacceptable.
\footnote{It is possible that radiative corrections with appropriate  soft SUSY breaking can induce additional terms \cite{maekawa} in the potential which can in principle allow SCPV but
this requires strong restrictions on soft $\chi$ mass terms.}
To allow SCPV, the minimal addition is of one singlet uncharged
scalar $\aleph$ which does not contribute to any anomaly and
allows  SCPV. This then completes the field content of our MSAM.

In the spirit of the aspon model we shall assume that the soft
breaking of SUSY respects CP invariance, {\it i.e.}
the lagrangian is of the form $\cal{L} = \cal{L}_{\it 0} +
\cal{L}_{S\!\!\!\!/U\!\!\!\!/S\!\!\!\!/Y\!\!\!\!/}$ and $\cal{L}$ is CP 
conserving. Recall that the quark mass matrices of the aspon 
model have the  texture:
\[ m_q = 
\left( \begin{array}{cc}
m & 0 \\
m_5 \alpha^{\dag} & M
\end{array} \right) \]
where $M$ denotes the mass of the vectorlike quark, $m_5$ the magnitude
of the mixing induced by  $\left\langle\chi_{1,2}\right\rangle$, and $\alpha$ the 
corresponding $3\times 1$ complex phase vector with $\alpha^{\dag} \alpha=1$. 
At tree-level $\bar{\theta}$ is zero. At one-loop order, both the gluino mass
and quark mass matrices develop imaginary parts and our main purpose 
is to find the constraints necessary to keep $\bar{\theta} < 10^{-9}$ as dictated by the bound on the neutron electric dipole moment.

The situation with SUSY is in this regard quite different from  the
non-supersymmetric aspon model where the one-loop
contribution\cite{aspon2} comes from only one specific diagram 
(Figure 1) which vanishes at lowest order in the minimal supersymmetric case. 
The reason is that the crucial four-scalar coupling $\lambda |\phi|^2|\chi|^2$ 
cannot arise from a superpotential which is gauge invariant. 
More important, there are now  new quark mass and  gluino mass one-loop 
diagrams (Figures 2 and 3 respectively) contributing significantly to
$\bar{\theta}$, as a result of soft SUSY breaking. For the 
quark mass diagram, the case with a gluino
running in the loop is by far the dominating one, due to the strong
QCD coupling. Cases involving other neutral gauginos with
the same structure are further suppressed.
Unlike the case analyzed in \cite{dine}, here we cannot
integrate out the vectorlike quark superfields before looking into the
constraints on the SSBP and have to consider mixings among both the 
left- and  right-handed quark states. This does  give constraints that
disappear in the large $M$ limit. In this sense, our treatment is 
complimentary to that in \cite{dine}. We first apply the bi-unitary
transformation diagonalizing the quark mass matrix to the superfields.
We write
\begin{equation}
U_R^{\dag} m_q U_L = 
 \left( \begin{array}{cc}m & 0 \\
0 & M \end{array} \right) 
\end{equation}
where, without loss of generality, we assumed the $3\times 3$ matrix
$m$ to be diagonal.
Then  the SSBP can be written in the form:
\begin{equation} 
U_R^{\dag} \widetilde{M}^2_{RL} U_L = \tilde{m}_{A}
 \left( \begin{array}{cc}m & 0 \\
0 & M \end{array} \right)
+ U_R^{\dag}  \left( \begin{array}{cc}\delta\! A \left\langle H \right\rangle & 0 \\
\delta\! A_5 m_5 a^{\dag} & \delta\! A_M M \end{array} \right) U_L  \; ; 
\end{equation}
\begin{equation} 
U_L^{\dag} \widetilde{M}^2_{LL} U_L =  \tilde{m}^2_{L} I
+ U_L^{\dag}  \left( \begin{array}{cc}\delta\!  \tilde{m}^2_{L} & 0 \\
0 & \delta\!  \tilde{M}^2_{L}\end{array} \right) U_L  \; ; 
\end{equation}
\begin{equation} 
U_R^{\dag} \widetilde{M}^2_{RR} U_R =  \tilde{m}^2_{R} I
+ U_R^{\dag}  \left( \begin{array}{cc}\delta\!  \tilde{m}^2_{R} & 0 \\
0 & \delta\!  \tilde{M}^2_{R}\end{array} \right) U_R  \; . 
\end{equation}
The quantities "$\delta\! A$" and "$\delta\! \tilde{m}^2$" parameterize
departure from proportionality and squark degeneracy.
The only possible {\it complex} quantities, arising from CP violating 
VEVs, among the 
SSBP are here absorbed into the $3\times 1$ phase vector
$a$, with $a^{\dag} a =1$. The second term in 
each of these expressions is in  general complex, as a result of the 
complex phases in $\alpha^{\dag}$ from the quark mass matrix going into the 
off-diagonal entries of $U_R$ and $U_L$. All the other parameters are
real. A $\delta\! A_M$ term can always be absorbed into $\tilde{m}_{A}$
but keeping it helps to illustrate some feature of the results below.  
The $\delta\! A_5$ term has a "hard" SUSY 
breaking piece involving VEV's of
the $F$-terms of the $\chi_{1,2}$ superfields,  $F_{\chi_{1,2}}$.
The exact definition for the term is given by
\begin{equation}
\delta\! A_5 m_5 a^{\dag}_i + \tilde{m}_{A}  m_5 \alpha^{\dag}_i
= A_{\chi_j}^i  h^j_i \left\langle \chi_j \right\rangle - h^j_i \left\langle F_{\chi_ j}\right\rangle
\end{equation}
where summation over $j=1$ and $2$ should be taken, and
$h^j_i$ denote Yukawa 
couplings  with then $m_5 \alpha^{\dag}_i = h^j_i  \left\langle \chi_j \right\rangle$. Alignment between
the phase vectors $\alpha^{\dag}$ and  $a^{\dag}$ is by no means guaranteed. This
is especially the case when the $\left\langle F_{\chi_j}\right\rangle$'s are nonzero.
However, if $\left\langle \chi_j \right\rangle$'s break $U(1)_a$ in a F-flat direction, 
misalignment between  $\alpha^{\dag}$ and  $a^{\dag}$ is then a direct
consequence of the lack of propotionality for the $A_{\chi_j}^i$'s.
\footnote{In principle, an alignment of the phases in $\left\langle F_{\chi_j}\right\rangle$'s
with those in $\left\langle \chi_j \right\rangle$'s would make the $F$-term contributions
themselves satisfy proportionality and be totally absorbed into the 
$\tilde{m}_{A}$ term. This alternative, however, does not seem to be
realistic.}
The above expressions (2-4)  clearly illustrate that in  the limit
of strict proportionality and degeneracy, the SSBP give {\it no}
contribution to $\bar{\theta}$.

 Now we go on to derive the constraints
on the lack of proportionality and degeneracy from  
$\bar{\theta} < 10^{-9}$. First note that both the  quark and
gluino mass diagrams require a $\widetilde{M}^2_{RL}$ mass
insertion on the squark line. The first order contribution to $\delta\! m_q$
(Fig. 2) is then proportional to $\widetilde{M}^2_{RL}$. However, as
the latter has the same type of texture as $m_q$, it always
gives a real trace to $m_q^{-1}\delta\! m_q$ and hence does not contribute 
to $\bar{\theta}$.  Similarly, if we take  only the proportional part of the
insertion and add an extra squark degeneracy violating mass insertion in
either the left- or the right-handed squark line, the trace is again real.
Complex phases in $m_q^{-1}\delta\! m_q$ contributing to $\bar{\theta}$
arise from:
{\it 1)} one proportionality violating insertion together with one degeneracy
violating insertion; {\it 2)} two degeneracy violating insertions;
{\it 3)} two proportionality violating insertions. For each
specific diagram of $\delta\! m_q$, there is a corresponding gluino mass
($\delta\! m_{\lambda}$) diagram (Fig.3) that is related  by interchanging
internal and external fermion lines. The $\delta\! m_{\lambda}$
diagram so obtained leads to a  $\bar{\theta}$ contribution suppressed
relative to the $\delta\! m_q$ diagram by $m_{i}^2/m_{\lambda}^2$ 
($m_i$ being the mass of a light quark) or $M^2/m_{\lambda}^2$ and
is hence uninteresting. The only exception is the case of a single
proportionality violating insertion; because unlike the $m_q^{-1}\delta\! m_q$
case where individual contributions to the imaginary part cancel in
the trace as noted above, here they give genuine contribution to
$\bar{\theta}$.  

To arrive at the explicit constraints, we use expressions  of  the $U_R$
and $U_L$ transformation matrices up to second order in 
$x$\cite{aspon2,x,aa} 
where $x=m_5/M$ is a small parameter characterizing mixing between
the light and heavy quarks. Actually, $x^2\sim 3\times 10^{-5}$ can
still give rise to sufficient CP violation in the $K-\bar{K}$ system
through the aspon exchange mechanism\cite{aspon2}. The strength of
each proportionality or degeneracy violating insertion can be obtained
by going to  the quark mass eigenstate basis and working out the second term
in each of the Eq.(2-4). To simplify the expressions, we use $\tilde{m}_S$
to denote the assumed common scale of  SSBP (including $m_{\lambda}$),
assuming also $M\mathrel{\raise.3ex\hbox{$<$\kern-.75em\lower1ex\hbox{$\sim$}}} \tilde{m}_S$. The constraints resulted are listed
in Table 1.

A few comments are in order. Firstly, the numerical constraints listed
in the Table are obtained by taking a "central" value for $x^2$
at $10^{-4}$ and assuming a common scale for the SSBP 
(including gaugino masses) with $M$ at about the same order. 
This choice of $x^2$ could possibly
be further reduced by up to an extra order of magnitude.
The smallness of the $x$ value is an important feature of the aspon
model that weakens the constraints on the SSBP, as compare to other 
generic SCPV schemes, and gives some hope that they can be satisfied. 
In specific SUSY breaking scenario with small 
$A$-terms\cite{gmsb,au1}, constraint expressions with a $\delta\! A/A$ factor
are explicitly weakened by an extra factor of $A/\tilde{m}_S$. Actually, small
$A$-term, e.g. $\sim 10^{-3} \tilde{m}_S$  goes a long way towards satisfying all constraints involving proportionality violations, except for constraint (6).  The only constraint not involving proportionality violation, no.(8), is also
much weakened due to a necessary $A$-insertion.
Reducing the gluino mass relative to squark masses strengthens the 
constraints (1)-(3) but weakens (4)-(10) by the same factor.
Second, we have taken $M\sim \tilde{m}_S$ which is what is to be
expected in the aspon model. The case of large $M$, however, cannot be
read off directly from the table. While constraint (7) is
reduced by at least a factor of $\tilde{m}_S^2/M^2$, others have to be
tracked down more carefully by identifying the heavy squark propagators
with masses $\sim (M^2+\tilde{m}_S^2)$ which are then dominated by
the supersymmetric contribution. When this is done carefully, the 
constraints  fall in agreement with the results in 
\cite{dine}. Note that to match our analysis with that of
 \cite{dine}, one has to take only the down-sector results and
flip the $L$ and $R$ indices. This leads to our last comment about
MSAM constraints.
We have been sticking to the original version of the aspon model
with a vectorlike quark doublet, which has constraints of the
form given in both the up- and down-sector. As illustrated in the Table,
some of the up-sector constraints are stronger than the corresponding
down-sector ones, essentially due to the heavy top mass.\footnote{
Further suppression to the down-sector 
constraints could be obtained in the large $\tan\beta$ setting, from 
$\left\langle H_d\right\rangle /\tilde{m}_S$.} In an alternative
aspon model with the vectorlike quark being a down-type 
singlet\footnote{Again, a flipping of $L$ and $R$ indices is needed.}, there
is no contribution to $\bar{\theta}$ from the up-sector and all 
constraints for the sector go away. This gives it an advantage. 
One should note that the vectorlike down-type singlet 
could {\it not} be replaced by an up-type one
--- CP violation then can only affect the $K-\bar{K}$ system through 
the KM-mechanism which requires $x\mathrel{\raise.3ex\hbox{$>$\kern-.75em\lower1ex\hbox{$\sim$}}} 0.1$. The essence of the aspon
model is then gone.\footnote{In relation to the
issue, an  interesting feature of the singlet version of the aspon model is
that CP violation in the up-sector is much suppressed relative to the down-sector, as the former has to come from the KM phase. This is
a unique characteristic of the model that could have interesting 
phenomenological consequences.} 
 
Now we take a brief look at what kind of features in a more complete model,
including details of the sector charged under $U(1)_a$ and a specific 
superysmmetry breaking mechanism, that have a better chance at
satisfying the constraints. The first thing we notice is that the doublet aspon
model is most probably unrealistic in a supersymmetric setting. The 
constraints  on up-type squark degeneracy are most certainly going to
be violated as a result of the large top Yukawa which enforces a much
larger renormalizaton group (RG-)running on the top squark, breaking
any degeneracy imposed at the SUSY breaking scale. For the 
rest of the discussion, we will concentrate on the aspon model with a
vectorlike down-type singlet. The degeneracy constraints are still stringent.
In particular there is one very strong constraint (no.(6)) requiring
\begin{equation}
\frac{\delta\! A}{A}
\frac{\delta\!  \tilde{m}^2_{L} - \delta\!  \tilde{M}^2_{L}}{\tilde{m}_S^2}
\mathrel{\raise.3ex\hbox{$<$\kern-.75em\lower1ex\hbox{$\sim$}}} 10^{-8}
\end{equation}
There is not much chance of satisfying the constraint together with all
the others without having 
$\frac{\delta\!  \tilde{m}^2_{R} - \delta\!  \tilde{M}^2_{R}}{\tilde{m}_S^2}
\mathrel{\raise.3ex\hbox{$<$\kern-.75em\lower1ex\hbox{$\sim$}}} 10^{-4}$ (remember that here $R$ reads $L$ from the table), at least. 
This sounds difficult. However,
once this condition is assumed, the other constraints on the SSBP
involving the scalar partner of the light quarks are in general {\it not}
much stronger than those demanded by FCNC experiments\cite{nc}.
This requires good degeneracy among the light ($\bar{d}$) singlets
and the new quark, $\bar{D}$. The latter  though having
the same standard model quantum number as the light singlets , 
bears an $U(1)_a$ charge. RG-running again 
distinguishes it from the light singlets with an effect dependent on the
the $U(1)_a$ gauge coupling. The best hope, we believe, is offered by
some sort of low-energy SUSY breaking scenario such as
gauge mediated models\cite{gmsb}. This tames the RG-runnings.
To name a possible scenario, if we have a gauge mediated model 
the messenger sector of which has no $U(1)_a$ charge, the effective
soft SUSY breaking terms would be blind to the  $U(1)_a$
as well as to flavor, thus allowing the degeneracy at first order. 

The more interesting part concerning our MSAM are the $\delta\! A_5$ related
constraints, as they are related to the $U(1)_a$ symmetry breaking.
Constraints (1) and (2) concern the lack of proportionality between
the $A_5$ term (c.f. Eq.5), coming from mixing between the 
light and heavy squarks, and the corresponding term in the superpotential.
Both terms are complex, as a result of SCPV coming with $U(1)_a$ 
symmetry breaking. What is needed is then a matching of the complex 
phases, $a^{\dag}\alpha \mathrel{\raise.3ex\hbox{$<$\kern-.75em\lower1ex\hbox{$\sim$}}} 10^{-3}$ from constraints (1). Constraint (2)
has a term with a slightly different phase structure but, for the 
down-sector only where the constraints  now apply, it goes away.
We have mentioned above that the symmetry
breaking has to go along a F-flat direction for $\chi_{1,2}$. Assuming
this, and  considering that the constraint (1) comes from the gluino 
mass diagram with a single proportionality violating insertion where 
each of the three families contribute independently, we can rewrite
the constraint as 
\begin{equation}
\delta\! A_{\chi}^i / A_{\chi}^i \mathrel{\raise.3ex\hbox{$<$\kern-.75em\lower1ex\hbox{$\sim$}}} 10^{-3}
\end{equation}
for each $i$ ($A_{\chi}^i \sim A_{\chi_j}^i \sim \tilde{m}_S$ is assumed).
This makes the physics content of the constraint more transparent.
Constraints  involving the
$\frac{\delta\! A_5 -\delta\! A_M}{\tilde{m}_S}$ factor imply a
more complicated restriction on the sector charged under $U(1)_a$.
The factor has phase vector components not shown explicitly in the Table. 
For example, the particularly stringent constraint (10) is actually
given by
\begin{equation}
\frac{\delta\! A}{A}
\frac{|\delta\! A_5 a_i^* -\delta\! A_M \alpha_i^*|}{\tilde{m}_S}
\mathrel{\raise.3ex\hbox{$<$\kern-.75em\lower1ex\hbox{$\sim$}}} 
10^{-8}\; .
\end{equation} 
In constraints (4) and (5)  the same factor, 
$\frac{|\delta\! A_5 a_i^* -\delta\! A_M \alpha_i^*|}{\tilde{m}_S}$ involved.
To suppress the factor requires alignment between the $A_{\chi}$-terms and 
the $A_M$ term, as well as the phase vectors $a$ and $\alpha$.

Of course we still have to write down a superpotential for the $\chi$ and
$\aleph$ superfields that breaks $U(1)_a$ in the way required. For
instance, we can have
\begin{equation}
W_{\chi,\aleph} = \sum_{j=1,2}^{k=3,4} y_{jk} \aleph \chi_j \chi_k + P(\aleph)
\end{equation}
where $ P(\aleph)$ is a general cubic polynomial in pure singlet 
$\aleph$. This is similar
to a well known example\cite{wb} from which one can easily see that
it admits a SUSY preserving vacuum with $\left\langle \aleph \right\rangle =0$.
This holds even in the presence of a Fayet-Iliopoulos $D$-term for
$U(1)_a$. 

In summary, the inclusion of low-energy supersymmetry makes it more
difficult to solve the strong CP problem with spontaneous CP violation.
We have constructed a minimal supersymmetric aspon model (MSAM)
with just one additional singlet superfield $\aleph$, and explicitly
evaluated the $\bar{\theta}$ constraints. The constraints on $A$-term
proportionality and squark degeneracy  require that 
the stringent inequality given by Eq.(6) be satisfied, but beyond that
the usual FCNC constraints for the MSSM are about sufficient. The major 
extra constraints are given by Eq.(7) and (8). It remains for future work to study 
whether the constraints can be satisfied in a more complete theory
incorporating specific mechanism of SUSY breaking.

We would like to acknowledge helpful discussions from A.W.Ackley,
R. Rohm and B.D.Wright. This work was supported in part by the U.S.
Department of Energy  under Grant No. DE-FG05-85ER-40219, Task B.

\clearpage

Table 1. Interesting  terms in $\bar{\theta}$ and estimates of resultant
constraints on the soft supersymmetry breaking parameters (SSBP). 
The magnitude of $\bar{\theta}$ contributions from each term has been split
into factors from different proportionality and squark degeneracy violations.
Explicit dependence on the phase vectors $a$ and $\alpha$ is shown only
for the first two constraints, where they may be of interest.
Numerical constraints listed in the last column, 
apply to the (products of) the  proportionality and/or degeneracy
violating  factor(s).  
No suppression from $\frac{\left\langle H\right\rangle}{\tilde{m}_S}$ is assumed.
$m_i$ could be the mass of a light quark of any family. 
A numerical 
constraint with  a $(u)$ or a $(d)$, refers to the up- or down-sector respectively;
otherwise, the constraint is common to both sectors. The constraints 
marked by * are the more important ones; all the others are very likely to
be satisfied if they are. 
%This is especially true for the down-sector
%constraints, the ones with interest for a possibly realistic
%minimal supersymmetric aspon model (MSAM).

\vspace*{.5in}
\begin{tabular}{c||c|ccc||c}\hline\hline
No. & \multicolumn{4}{c||}{ magnitude of $\bar{\theta}$ contribution} & constraints \\ \hline
(1)$^*$ & $\frac{3 \alpha_s}{4\pi} x^2 \frac{M^2}{\tilde{m}_S^2}$ &
$\frac{\delta\! A_5  Im(a^{\dag}\alpha)}{\tilde{m}_S}$ & & &
$\mathrel{\raise.3ex\hbox{$<$\kern-.75em\lower1ex\hbox{$\sim$}}} 10^{-3}$ \\
(2) & $\frac{3 \alpha_s}{4\pi} x^2$ &
$ \frac{\delta\! A_5 ImTr (a^{\dag}m^2 \alpha)}{\tilde{m}_S^3}$ & & &
$\mathrel{\raise.3ex\hbox{$<$\kern-.75em\lower1ex\hbox{$\sim$}}} 10^{-3}$ \\
(3) & $\frac{3 \alpha_s}{4\pi} x^2$ 
$\frac{\left\langle H\right\rangle  m_i}{\tilde{m}_S^2}$ &
$\frac{\delta\! A}{A}$ & & & $\surd$ by (9)\\
%$\mathrel{\raise.3ex\hbox{$<$\kern-.75em\lower1ex\hbox{$\sim$}}} %10^{-1}_{\ \ (d)} \; ; 10^{-3}_{\ \ (u)}$\\
(4) & $\frac{\alpha_s}{4\pi} x^2$ &
$\frac{\delta\! A_5 -\delta\! A_M}{\tilde{m}_S}$ &
$\frac{\delta\!  \tilde{m}^2_{L} - \delta\!  \tilde{M}^2_{L}}{\tilde{m}_S^2}$ & & 
$\mathrel{\raise.3ex\hbox{$<$\kern-.75em\lower1ex\hbox{$\sim$}}} 10^{-3}$\\
(5) & $\frac{\alpha_s}{4\pi} x^2$ &
$\frac{\delta\! A_5 -\delta\! A_M}{\tilde{m}_S}$ & &
$\frac{\delta\!  \tilde{m}^2_{R} - \delta\!  \tilde{M}^2_{R}}{\tilde{m}_S^2}$  & $\mathrel{\raise.3ex\hbox{$<$\kern-.75em\lower1ex\hbox{$\sim$}}} 10^{-3}$\\
(6)$^*$ & $\frac{\alpha_s}{4\pi} x^2 \frac{\left\langle H\right\rangle}{m_i}$ &
$\frac{\delta\! A}{A}$ & 
$\frac{\delta\!  \tilde{m}^2_{L} - \delta\!  \tilde{M}^2_{L}}{\tilde{m}_S^2}$ & &
$\mathrel{\raise.3ex\hbox{$<$\kern-.75em\lower1ex\hbox{$\sim$}}} 10^{-8}$\\
(7) & $\frac{\alpha_s}{4\pi} x^2 \frac{\left\langle H\right\rangle m_i}{M^2}$ &
$\frac{\delta\! A}{A}$ & &
$\frac{\delta\!  \tilde{m}^2_{R} - \delta\!  \tilde{M}^2_{R}}{\tilde{m}_S^2}$ & 
$\surd$ by (9)\\
(8)$^*$ & $\frac{\alpha_s}{4\pi} x^2 \frac{A}{\tilde{m}_S}$ 
$\frac{m_i}{m_j}\ (i\neq j)$ & &
$\frac{(\delta\!  \tilde{m}^2_{L}, \delta\!  \tilde{M}^2_{L})}{\tilde{m}_S^2}$ &
$\frac{\delta\!  \tilde{m}^2_{R}}{\tilde{m}_S^2}$ & 
$\mathrel{\raise.3ex\hbox{$<$\kern-.75em\lower1ex\hbox{$\sim$}}} 10^{-6}_{\ \ (d)} \; ; 10^{-7}_{\ \ (u)}$\\
(9)$^*$ &  $\frac{\alpha_s}{4\pi} x^2
\frac{A  \left\langle H\right\rangle  ^2}{\tilde{m}_S^3}$ 
$\frac{m_i}{m_j}\ (i\neq j)$ & $\left(\frac{\delta\! A}{A}\right)^2$ & & &
$\mathrel{\raise.3ex\hbox{$<$\kern-.75em\lower1ex\hbox{$\sim$}}} 10^{-6}_{\ \ (d)} \; ; 10^{-7}_{\ \ (u)}$\\ 
(10)$^*$ &  $\frac{\alpha_s}{4\pi} x^2
\frac{A \left\langle H\right\rangle M^2}{\tilde{m}_S^3 m_i}
$ & $\frac{\delta\! A}{A}
\frac{\delta\! A_5 -\delta\! A_M}{\tilde{m}_S}$ & & &
$\mathrel{\raise.3ex\hbox{$<$\kern-.75em\lower1ex\hbox{$\sim$}}} 
10^{-8}$\\
\hline\hline
\end{tabular}

\newpage

\bigskip

{\bf Figure Captions:-}

\medskip
\medskip

{\bf 1.} 1-loop quark mass diagram contributing to  $\bar{\theta}$  in the nonsupersymmetric aspon model (with vectorlike doublet $Q_o$).\\

\medskip

{\bf 2.} 1-loop quark mass diagram  contributing to  $\bar{\theta}$  in the supersymmetric aspon model.\\

\medskip

{\bf 3.} 1-loop gluino mass diagram  contributing to  $\bar{\theta}$  in  the supersymmetric aspon model.\\

\clearpage

\begin{figure}
\centering
\hspace{0in}\epsffile{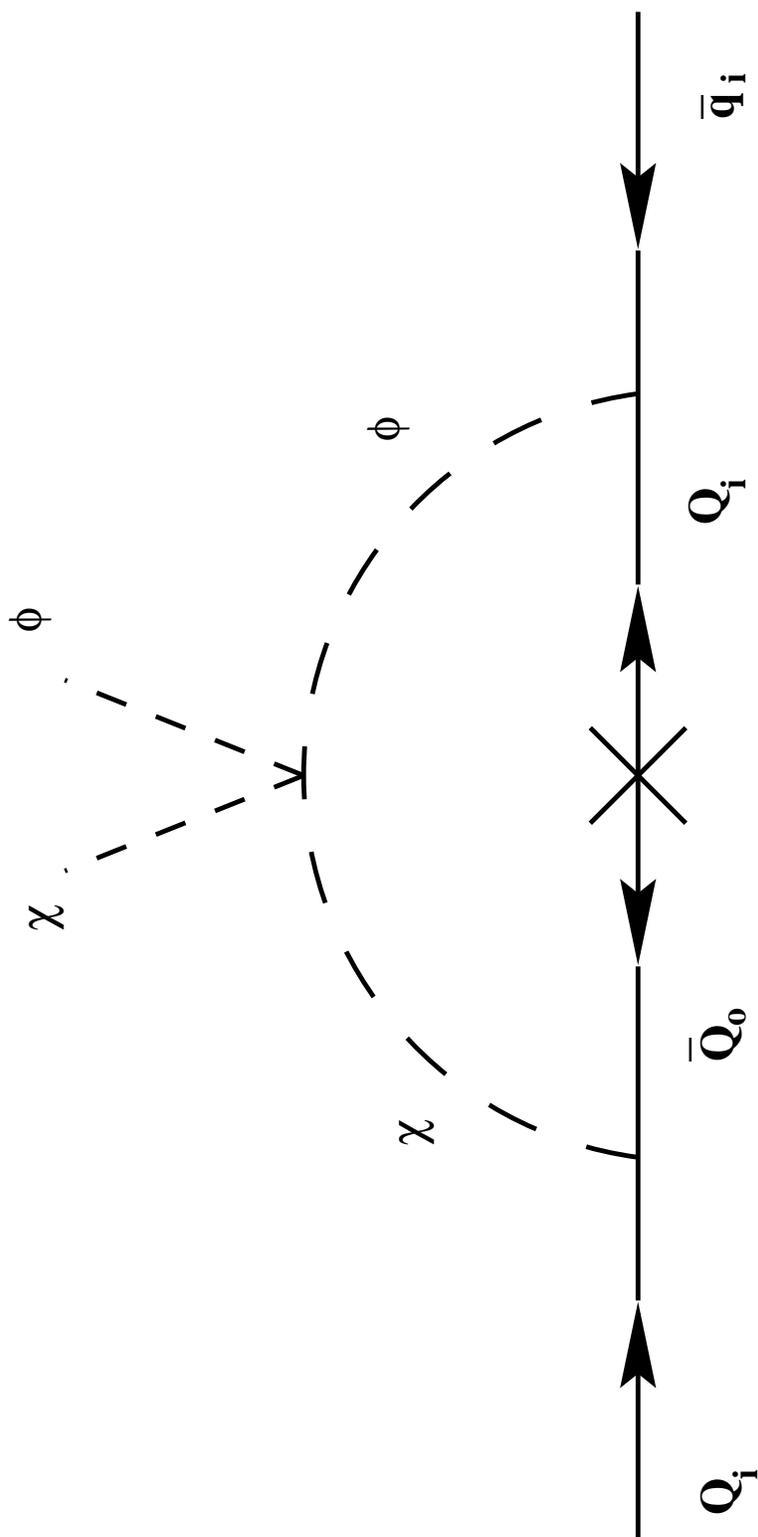}

\caption{1-loop quark mass diagram contributing to  $\bar{\theta}$  in the nonsupersymmetric aspon model (with vectorlike doublet $Q_o$).}
\end{figure}

\clearpage

\begin{figure}
\centering
\hspace{0in}\epsffile{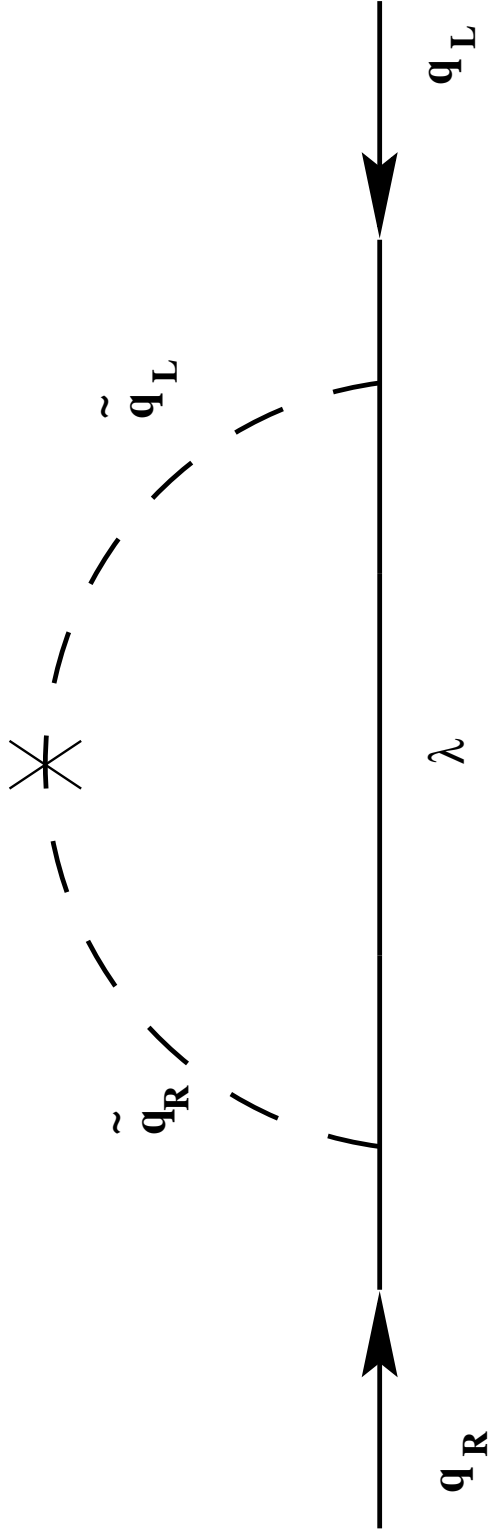}

\caption{ 1-loop quark mass diagram  contributing to  $\bar{\theta}$  in the supersymmetric aspon model.}
\end{figure}

\clearpage

\begin{figure}
\centering
\hspace{0in}\epsffile{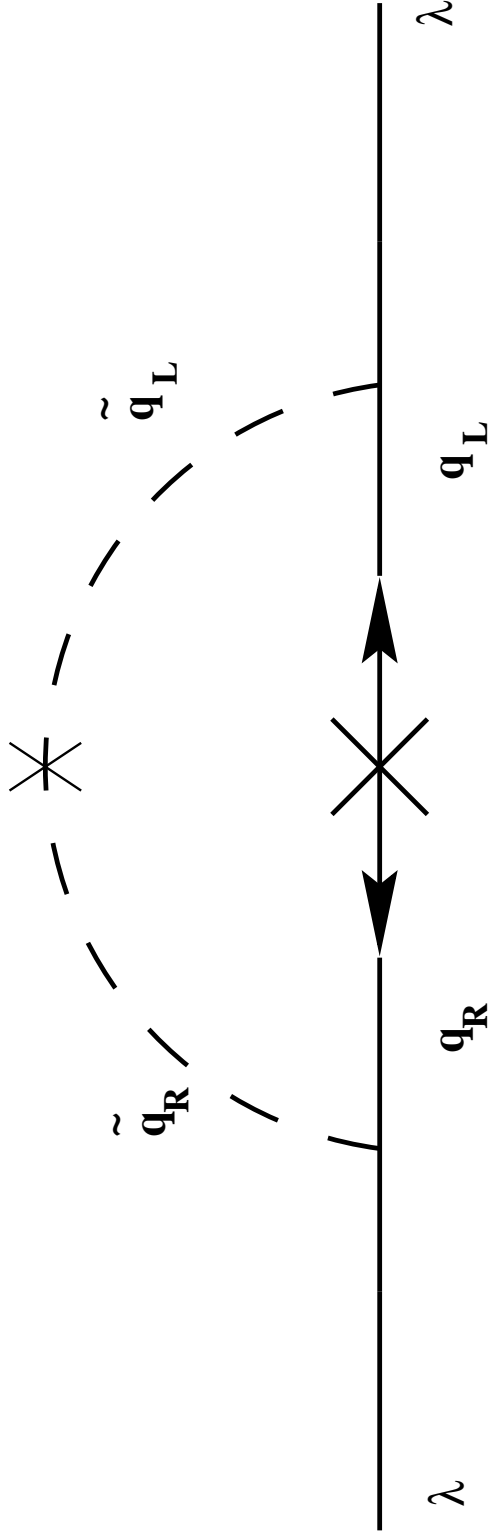}

\caption{1-loop gluino mass diagram  contributing to  $\bar{\theta}$  in  the supersymmetric aspon model.}
\end{figure}

\end{document}